 \documentclass[aip,pop,reprint]{revtex4-1}
\usepackage{dcolumn}
\usepackage{bm}
\usepackage{amsmath}
\usepackage{tabularx}

\newcommand{\bb}{\textrm{B}}

\newcommand{\vc}{\mathbf}
\newcommand{\ext}{\textrm{ext}}
\def\be{\begin{eqnarray}}
\def\ee{\end{eqnarray}}
\def\ben{\begin{eqnarray*}}
\def\een{\end{eqnarray*}}

\usepackage{graphicx}
\usepackage{color}

\begin{document}


\title{Mean Force Kinetic Theory:
a Convergent Kinetic Theory for Weakly and Strongly Coupled Plasmas}


\author{Scott D.\ Baalrud}

\affiliation{Department of Physics and Astronomy, University of Iowa, Iowa City, IA 52242}
\author{J\'{e}r\^{o}me Daligault}
\affiliation{Theoretical Division, Los Alamos National Laboratory, Los Alamos, New Mexico 87545}


\date{\today}

\begin{abstract}
A new closure of the BBGKY hierarchy is developed, which results in a convergent kinetic equation that provides a rigorous extension of plasma kinetic theory into the regime of strong Coulomb coupling.
The approach is based on a single expansion parameter which enforces that the exact equilibrium limit is maintained at all orders. 
Because the expansion parameter does not explicitly depend on the range or the strength of the interaction potential, the resulting kinetic theory does not suffer from the typical divergences at short and long length scales encountered when applying the standard kinetic equations to Coulomb interactions.
The approach demonstrates that particles effectively interact via the potential of mean force and that the range of this force determines the size of the collision volume. 
When applied to a plasma, the collision operator is shown to be related to the effective potential theory [Baalrud and Daligault, Phys.~Rev.~Lett {\bf 110}, 235001 (2013)].
The relationship between this and previous kinetic theories is discussed. 
\end{abstract}




\maketitle


\section{Introduction\label{sec:intro}}

Derivations of standard plasma kinetic equations confront either infrared divergences (Boltzmann equation)~\cite{grad:58,harr:71,ferz:72},  ultraviolet divergences (Lenard-Balescu equation)~\cite{lena:60,bale:60,guer:60}, or both (Landau equation)~\cite{land:36}. 
These divergences are resolved by invoking physical arguments, such as Debye shielding to resolve the infrared divergence, or the distance of closest approach in a binary collision to resolve the ultraviolet divergence. 
The results obtained in either approach agree to logarithmic accuracy as long as the plasma is weakly coupled and near local equilibrium.
They have been validated for many collisional transport processes satisfying these limits over the several decades since they were developed.
Despite the success of these equations, the theory stands in an unsatisfactory position in that it relies on ad hoc arguments. 
As a result, it is difficult to attempt generalizations to other important situations, such as moderate or strong Coulomb coupling, or strong magnetization.

Using a new self-consistent closure of the BBGKY hierarchy, we present a plasma kinetic equation that (i) is convergent, i.e. it does not confront either infrared or ultraviolet divergences, and that (ii) provides a rigorous extension of plasma kinetic theory into the regime of strong Coulomb coupling.
The approach demonstrates that particles effectively interact via the potential of mean force and that the range of this force determines the size of the collision volume. 

In order to put our approach into perspective, we first briefly recall the origin of standard kinetic equations.
Each can be derived from a perturbative closure of the BBGKY hierarchy obtained by first identifying a small dimensionless parameter characteristic of the system, and by then using its smallness to truncate the hierarchy at the level of two-particle correlations.
Two dimensionless parameters naturally arise in terms of the average particle density $n$, and the range $l$ and the strength $\phi_0$ of the interaction potential $\phi(r)$.
Namely, the ``concentration parameter'' $nl^3$,  which measures how many particles, on the average, simultaneously interact with a given
particle, and the ``strength parameter'' $\phi_0/k_\bb T$, which measures the interaction potential energy of two colliding particles in units of the mean particle kinetic energy.
It is easily seen that, strictly speaking, these parameters are not well adapted to plasmas since the Coulomb potential has an infinite range, and its magnitude becomes arbitrarily strong at short distance.
The celebrated Boltzmann equation is obtained by assuming $nl^3\ll1$ and $\phi_0/k_\bb T\sim O(1)$; it applies to dilute gases interacting via short range potentials of arbitrary interaction strength.~\cite{grad:58,harr:71,ferz:72} 
The Landau equation is obtained in the so-called weakly coupled limit characterized by $\phi_0/k_\bb T\ll 1$ and $nl^3\sim O(1)$; it applies to weak potentials, i.e. potentials that are uniformly small for all interparticle distances $r$, a condition that is generally not respected due to the typical strong repulsions at short distances.~\cite{land:36}
The Lenard-Balescu equation, which was developed to deal with weakly coupled plasmas, is obtained in the so-called weakly-coupled, long-range limit assuming $\phi_0/k_\bb T\ll 1$ and $nl^3 \phi_0/k_\bb T\sim O(1)$.~\cite{lena:60,bale:60,guer:60} 
Plasmas do not fall in any of these categories, which explains why one needs to regularize {\it ad-hoc} the standard kinetic equations before applying them to plasmas.
A solution to the problem can be obtained by writing the Coulomb potential as a sum of two terms: a weak long-range term, plus a strong short-range term. 
This amounts to carefully joining the Lenard-Balescu and Boltzmann equations (see, e.g., Refs.~\onlinecite{thom:60,hubb:61,bald:62,frie:63,kiha:63,wein:64,aono:65,goul:67}).

Until today, no one has derived a {\it practical} kinetic equation beyond the previous limits.
The generalization of the Boltzmann equation to higher densities remains an unsolved challenge (much progress on the kinetic theory of dense gases and liquids has been made, but for systems in thermal equilibrium).
It was shown that the systematic inclusion of many-body collisions through a density expansion similar to the virial expansion for computing the equilibrium properties of dense gases is plagued by unphysical divergences.\cite{dorf:67}
This is because $N$-body collisions and correlations cannot be treated separately from those of all higher orders ($N\!+\!1$-body, $N\!+\!2$-body, etc.).
This difficulty is symptomatic of many perturbative approaches in physics and could in principle be dealt with using so-called renormalization techniques.
While such renormalization schemes have been proposed,\cite{dali:11} their exceeding complexity has thus far restricted their usefulness to formal, but not practical, solutions.

In this paper, we present a closure scheme for the BBGKY hierarchy that, like the standard equations, results from the identification of a small expansion parameter but, unlike the concentration and strength parameters, does not explicitly depend on the characteristics of the interaction potential like $l$ and $\phi_0$.
The present expansion parameter [Eq.~(\ref{eq:dfn}) below] is a measure of the perturbation of the distribution function about thermal equilibrium.
The closure retains the correct equilibrium limit at all orders, and ensures that screening is captured in the near-equilibrium limit, while at the same time allowing inclusion of short-range interactions. 
It results in a convergent and tractable kinetic equation that provides a rigorous extension of plasma kinetic theory into the regime of strong Coulomb coupling.
Here, strong Coulomb coupling refers to plasmas in which the average
Coulomb potential energy of interacting particles
$(Ze)^2/4\pi\epsilon_0 a$ ($Ze$ is the electric charge, $a=(3/4\pi n)^{1/3}$ is the average interparticle spacing) exceeds their average kinetic energy $k_\bb T$ ($T$ is the temperature), i.e. $\Gamma \gtrsim 1$, where
\begin{eqnarray}
\Gamma = \frac{(Ze)^2/4\pi\epsilon_0 a}{k_\bb T} \,.\label{eq:gamma}
\end{eqnarray}

The resulting collision operator is the same as what is obtained from the Boltzmann equation if the interparticle force in binary collisions is taken to be the potential of mean force,~\cite{hill:60} rather than the Coulomb potential. 
In the weakly coupled limit, the potential of mean force is the Debye-H\"{u}ckel potential~\cite{deby:23} and the theory gives the same results as standard plasma kinetic equations.~\cite{baal:14}
The transport coefficients resulting from the convergent kinetic equation are those of the recent ``effective potential theory'' (EPT), which was shown to successfully extend the conventional plasma transport theory into the strongly coupled regime.\cite{baal:13,baal:14,baal:15}
Previous work has shown that EPT can extend plasma kinetic theory well into the regime of strong Coulomb coupling.~\cite{dali:14,bezn:14,haxh:14,stri:16,dali:16,shaf:17} 

In addition, in the resulting hydrodynamic equations the pressure and internal energy are composed of two parts; a kinetic (or ideal) part and a potential (or excess) part.
The ideal part, which is obtained with all kinetic equations, is the standard ideal gas component that represents the transfer of momentum or energy due to the flow of particles.
The excess part, which is typically absent from standard kinetic theories, represents the transfer of momentum or energy between particles by the particle interactions.

The paper is organized as follows.
In Section~\ref{section_II}, we introduce the expansion parameter at the basis of the closure of the BBGKY hierarchy.
In Section~\ref{sec:c_op}, the convergent kinetic equation resulting from this closure is derived and its properties are discussed.
To this end, we closely follow the method used by Grad in his derivation of the Boltzmann equation.~\cite{grad:58} 
In Section~\ref{sec:comp}, the convergent kinetic equation is discussed in comparison to the standard plasma kinetic equations.

\section{Basic expansion parameter} \label{section_II}

Kinetic theories can be derived from the BBGKY hierarchy~\cite{ferz:72} 
\begin{align}
\label{eq:bbgky}
\biggl[ \frac{\partial}{\partial t}  &+ \sum_{i=1}^n \biggl( \mathcal{L}_i +\mathcal{L}^{\ext}_i+ \sum_{\stackrel{j=1}{j\neq i}}^n \mathcal{L}_{ij}^C \biggr) \biggr] f^{(n)}(\vc{r}^n, \vc{v}^n,t) \\ \nonumber
&= -\sum_{i=1}^n \int d\vc{\Gamma}_{n+1} \mathcal{L}^C_{i,n+1}  f^{(n+1)}(\vc{r}^{n+1}, \vc{v}^{n+1},t)  ,
\end{align}
where 
\begin{equation}
f^{(n)}(\vc{r}^n, \vc{v}^n,t) = \frac{N!}{(N-n)!} \int d\vc{\Gamma}^{(N-n)} f^{[N]} (\vc{r}^N, \vc{v}^N,t) 
\end{equation}
defines the $n^{\textrm{th}}$-order reduced distribution functions in terms of the $N$-particle distribution function $f^{[N]}$. 
Here,  $\vc{r}^N = (\vc{r}_1, \vc{r}_2, \ldots \vc{r}_N)$, where $\vc{r}_i$ denotes the spatial location of particle $i$, $\vc{v}^N = (\vc{v}_1, \vc{v}_2, \ldots \vc{v}_N)$ where $\vc{v}_i$ denotes the velocity of particle $i$, $d\vc{\Gamma}_n \equiv d\vc{r}_n d\vc{v}_n$ is a shorthand notation for the 6-dimensional phase-space, and $d\vc{\Gamma}^{(N-n)} \equiv d\vc{\Gamma}_{n+1} \ldots d\vc{\Gamma}_{N}$. 
The BBGKY hierarchy follows from integrating the Liouville equation
\begin{equation}
\label{eq:liou}
\biggl[ \frac{\partial}{\partial t} + \sum_{i=1}^N \biggl( \mathcal{L}_i +\mathcal{L}^{\ext}_i+ \sum_{\stackrel{j=1}{j\neq i}}^N \mathcal{L}_{ij}^C \biggr) \biggr] f^{[N]} (\vc{r}^N, \vc{v}^N ,t) = 0 ,
\end{equation}
in order to obtain an evolution equation for each reduced distribution function. 
In equation (\ref{eq:bbgky}),
\begin{eqnarray}
\mathcal{L}_i = \vc{v}_i \cdot \frac{\partial}{\partial\vc{r}_i}+ \frac{q_i}{m_i} (\vc{v}_i \times \vc{B}) \cdot \frac{\partial}{\partial \vc{v}_i}
\end{eqnarray}
where the second term is associated with the Lorentz force due to an external magnetic field, and
\begin{eqnarray}
\mathcal{L}_i^{\ext}=\frac{1}{m_i}\vc{F}_i^{\ext} \cdot \frac{\partial}{\partial \vc{v}_i}
\end{eqnarray}
where $\vc{F}_i^{\ext} = -\nabla_i \phi_{\ext}(\vc{r}_i)$ is an external force, associated with an external potential. 
The external potential is added here for later convenience to introduce spatial inhomogeneities in the thermal equilibrium state.
The term
\begin{equation}
\mathcal{L}^C_{ij} \equiv \frac{1}{m_i} \vc{F}^C_{ij} \cdot \frac{\partial}{\partial \vc{v}_i}
\end{equation}
is associated with the  electrostatic Coulomb interactions between particles 
\begin{equation}
\vc{F}^C_{ij} = \frac{q_i q_j}{4\pi \epsilon_o } \frac{(\vc{r}_i - \vc{r}_j)}{|\vc{r}_i - \vc{r}_j|^3}  = - \nabla_i \phi(r_{ij})  .
\end{equation}

Any kinetic theory aims to describe the evolution of the first-order reduced distribution function $f^{(1)}(\vc{r}, \vc{v},t)$ ($n=1$):
\begin{equation}
\label{eq:n1}
\biggl( \frac{\partial }{\partial t} + \mathcal{L}_1 +\mathcal{L}^{\ext}_1\biggr) f^{(1)}(1)  = - \int d\vc{\Gamma}_2\, \mathcal{L}_{12}^C f^{(2)}(1,2) ,
\end{equation}
which depends on $f^{(2)}(1,2)$ [$(1,2)$ is shorthand notation for $(\vc{r}_1, \vc{v}_1, \vc{r}_2, \vc{v}_2, t)$]. 
The task of a collision operator is to provide an approximate expression for $f^{(2)}(1,2,t)$. 
For example, this can be related to $f^{(3)}(1,2,3,t)$ via the $n=2$ equation 
\begin{align}
\label{eq:f2}
\biggl( \frac{\partial}{\partial t} &+ \mathcal{L}_{1} + \mathcal{L}^{\ext}_1+\mathcal{L}_2 +\mathcal{L}^{\ext}_2+ \mathcal{L}_{12}^C + \mathcal{L}_{21}^C \bigg) f^{(2)}(1,2) \\ \nonumber
&= - \int d\vc{\Gamma}_3 \biggl( \mathcal{L}_{13}^C + \mathcal{L}_{23}^C \biggr) f^{(3)}(1,2,3)   ,
\end{align}
but approximations must be introduced to close the hierarchy. 

The most famous closure is to neglect the third order
distribution function $f^{(3)} \rightarrow 0$ in Eq.~(\ref{eq:f2}), 
which is an excellent approximation for dilute gasses that interact
via a short-range force (in the notation of the introduction, $nl^3\ll
1$). 
One then solves the homogeneous equation for $f^{(2)}$ subject to a constraint on the length scale over which binary collisions occur, as well as a lack of initial correlations in a binary scattering event.  
This method was shown to lead to the Boltzmann equation.
The calculation is not trivial and several variations 
exist; below we shall rely on the method proposed by Grad.~\cite{grad:58}
However, as recalled in the introduction, this closure is a poor approximation for plasmas. 
The reason is that the physics of screening is contained in $f^{(3)}$. 
Neglecting screening not only misses an important physical process, but the resulting kinetic equation diverges because the two-body Coulomb force has an infinite range. 


The closure scheme proposed here enforces that the correct equilibrium (i.e., thermodynamic) limit is maintained at all orders. 
This ensures that screening is captured, while at the same time allowing inclusion of short-range interactions. 
This can be accomplished by taking the basic expansion parameter to be  
\begin{equation}
\label{eq:dfn}
\Delta f^{(n+1)} \equiv  f_o^{(n+1)} \biggl( \frac{f^{(n+1)}}{f_o^{(n+1)}} - \frac{f^{(n)}}{f_o^{(n)}} \biggr).
\end{equation}
Equation~(\ref{eq:dfn}) measures the difference of the $n+1$ and $n$ probability distributions, referenced to their equilibrium values. 
It is a measure of the perturbation of non-equilibrium correlations about equilibrium; $\Delta f^{(n+1)}\rightarrow 0$ represents that correlations approach their value at equilibrium. 
In Eq.~(\ref{eq:dfn}),
\begin{equation}
f_o^{(n)}(\vc{r}^n, \vc{v}^n) = \rho^{(n)} (\vc{r}^n) f_\textrm{M}^{(n)}(\vc{v}^n)
\end{equation}
is the equilibrium reduced distribution function, 
\begin{equation}
f_\textrm{M}^{(n)} (\vc{v}^n) = \biggl(\frac{m}{2\pi k_\bb T}\biggr)^{3n/2} \exp \biggl( - \sum_{i=1}^n \frac{m \vc{v}_i^2}{2k_\bb T} \biggr)
\end{equation} 
is the Maxwellian velocity distribution function, 
\begin{equation} 
\rho^{(n)} (\vc{r}^n) = \frac{N!}{(N-n)!} \frac{1}{\mathcal{Z}_N} \int d\vc{r}^{(N-n)}  e^{-(V_{\ext}+V_N)/k_\bb T} \label{rhonrn}
\end{equation} 
is the $n-$particle density distribution function, $V_N(\vc{r}^N) = \sum_{i=1}^N \sum_{j>i}^N \phi(r_{ij})$ is the electrostatic potential energy, $V_{\ext}(\vc{r}^N)=\sum_{i=1}^N{\phi_{\ext}(\vc{r}_i)}$ is the interaction energy with the external potential $\phi_{\ext}$, and $\mathcal{Z}_N = \int \exp\left[-(V_{\ext}+V_N)/k_\bb T\right] d\vc{r}^N$ is the configurational integral. 

The BBGKY hierarchy from Eq.~(\ref{eq:bbgky}) can be rearranged in a way that expresses $\Delta f^{(n+1)}$ as an expansion parameter
\begin{equation}
\label{eq:bbgky_2}
\biggl[ \frac{\partial}{\partial t} + \sum_{i=1}^n (\mathcal{L}_i + \bar{\mathcal{L}}_i^{(n)} ) \biggr] f^{(n)} =
- \sum_{i=1}^n \int d\vc{\Gamma}_{n+1} \mathcal{L}_{i,n+1}^C \Delta f^{(n+1)} ,
\end{equation}
where 
\begin{subequations}
\begin{align}
\bar{\mathcal{L}}_i^{(n)} &\equiv \frac{1}{m_i} \bar{\vc{F}}_i^{(n)}(\vc{r}^n) \cdot \frac{\partial}{\partial \vc{v}_i} \\ 
&= \mathcal{L}^{\ext}_i+\sum_{\stackrel{j=1}{j\neq i}}^n \mathcal{L}_{ij}^C + \int d\vc{\Gamma}_{n+1} \mathcal{L}_{i,n+1}^C \frac{f_o^{(n+1)}}{f_o^{(n)}}
\end{align} 
\end{subequations}
is an operator associated with force 
\begin{eqnarray}
\label{eq:Fmf}
\bar{\vc{F}}_i^{(n)}(\vc{r}^n) &=& \vc{F}_i^{\ext} + \sum_{\stackrel{j=1}{j\neq i}}^n \vc{F}^C_{ij}\\
&+& \int \vc{F}^C_{i,n+1} \frac{\rho^{(n+1)} (\vc{r}^{n+1})}{\rho^{(n)} (\vc{r}^n)} d\vc{r}_{n+1} .\nonumber
\end{eqnarray}
The force (\ref{eq:Fmf}) has a simple physical interpretation. 
It is the mean force acting on particle $i$ obtained when keeping both it and a set of other particles ($j=1,\ldots ,n$, excluding $j=i$) at fixed positions and averaging over all equilibrium configurations of the other $N-n$ particles. 
As shown in Appendix~\ref{pomf_appendix}, this statistical force can be expressed as the gradient of a potential,
\begin{equation}
\label{eq:fmf}
\bar{\vc{F}}_i^{(n)} (\vc{r}^n) = - \nabla_i w^{(n)} (\vc{r}^n) ,
\end{equation}
where the potential of mean force is 
\begin{equation}
\label{eq:pomf}
w^{(n)}(\vc{r}^n) = - k_\bb T \ln \left[\frac{\rho^{(n)}(\vc{r}^{n})}{\rho^n}\right]\,,
\end{equation}
where $\rho=\rm lim_{V\to\infty}{\int_V{d\vc{r}\rho^{(1)}({\bf r})}/V}$ is the average particle density.
In the absence of the external potential, $\phi_{\ext}\equiv 0$, $w^{(n)}(\vc{r}^n)=-k_\bb T\ln g^{(n)}(\vc{r}^n)$, where 
$g^{(n)}(\vc{r}^n) = \rho^{(n)}(\vc{r}^n)/\rho^n$ is the $n$-particle distribution function (when $\phi_{\ext}\neq 0$, 
$g^{(n)}(\vc{r}^n) = \rho^{(n)} (\vc{r}^n)/\Pi_{i=1}^n \rho^{(1)}(\vc{r}_i)$). 

Although Eq.~(\ref{eq:bbgky_2}) is equivalent to Eq.~(\ref{eq:bbgky}), writing it in this way affords certain pedagogical clarities. 
The right-hand side, at any order $n$, can now be interpreted as a ``collision operator'' in the sense that it vanishes at equilibrium and is small compared to the left side of the equation for slight perturbations from equilibrium. 
This contrasts with the right side of Eq.~(\ref{eq:bbgky}), which does not vanish at equilibrium. 
Of course, solving Eq.~(\ref{eq:bbgky_2}) still requires a closure. 
The scheme suggested here is that the dynamical evolution of $f^{(n)}$ be closed at order $n$ by taking $\Delta f^{(n+1)} \rightarrow 0$. 
However, a closure for the equilibrium distribution $\rho^{(n)}(\vc{r}^n)$ (or $g^{(n)}(\vc{r}^n)$ in the homogeneous case) is still required in order to determine the potential of mean force arising on the left side of the equation. 
Determining $\rho^{(n)}$ is a more tractable problem because one can rely on methods of equilibrium statistical mechanics. 
At equilibrium, the BBGKY hierarchy reduces to the Yvon-Born-Green (YBG) limit~\cite{hans:06}
\begin{align}
\label{eq:ybg}
\nabla_1 \rho^{(n)}(\vc{r}^n) &- \frac{1}{k_\bb T} \biggl( \vc{F}_1^{\ext} + \sum_{j=2}^n \vc{F}_{1,j}^C \biggr) \rho^{(n)}(\vc{r}^n) \\ \nonumber
&= \frac{\rho}{k_\bb T} \int d\vc{r}_{n+1} \vc{F}_{1,n+1} \rho^{(n+1)}(\vc{r}^{n+1}) .
\end{align} 
Although this is also a hierarchical equation, accurate approximations have been developed for most potentials of interest and for quite general conditions of density and temperature.~\cite{hans:06} 
We also note that Eq.~(\ref{eq:pomf}) follows directly from Eq.~(\ref{eq:ybg}). 
To proceed with the kinetic theory derivation, $\rho^{(n)}$ will be considered a known quantity. 

In particular, in the following section a kinetic equation is derived from the second order ($n=2$) term of Eq.~(\ref{eq:bbgky_2}) assuming that $\Delta f^{(3)}=0$. 
The derivation closely follows Grad's method.~\cite{grad:58}
Many methods have been used to derive the Boltzmann equation from~Eq.~(\ref{eq:bbgky}),~\cite{ferz:72,cohe:61} and any of these could also be used for our purposes.
We choose Grad's method because it follows directly from the closure $f^{(3)} \rightarrow 0$ (or $\Delta f^{(3)}\rightarrow 0$ in our modification) without the need to introduce additional complications associated with a cluster expansion. 
It will also give a clear description of the effective collision volume in a Coulomb system. 
The closure for $\rho^{(2)}(\vc{r}_1, \vc{r}_2)$ will be assumed known. 
We will focus on the near-homogenous limit that is relevant to fluid theory. 
Here, the two-particle density consists of a component that varies on large scales indicative of the fluid-scale gradients, and another short spatial scale indicative of the scale of interactions: $\rho^{(2)} \approx \rho (\vc{r}_1) g^{(2)}(\vc{r}_1, r)$, where $r = |\vc{r}_1 - \vc{r}_2|$. 
The radial distribution function $g^{(2)}(r)$ can be provided by the hypernetted chain approximation,~\cite{hans:06} which is known to be accurate for plasmas at conditions spanning weak to strong coupling. 
However, the theory does not depend on the method used to obtain $g^{(2)}$. 
At weak coupling, $g^{(2)}$ simply asymptotes to the Debye-H\"{u}ckel limit and conventional plasma kinetic theory will result from this unified framework. 

\section{Plasma Kinetic Equation\label{sec:c_op}} 

\subsection{Modified Grad Method}

The plasma kinetic equation (\ref{eq:n1}) in terms of the expansion parameter of Eq.~(\ref{eq:dfn}) is 
\begin{equation}
\label{eq:n1_f}
\biggl( \frac{\partial}{\partial t} + \mathcal{L}_1 + \bar{\mathcal{L}}_1^{(1)} \biggr) f^{(1)}(1) = - \int d\vc{\Gamma}_2 \mathcal{L}_{12}^{C} \Delta f^{(2)}(1,2) .
\end{equation}
Following Grad's method,~\cite{grad:58} space is divided into a volume within which collisions occur $V_\sigma$, and a volume outside of this in which the observable (truncated) distribution functions are defined 
\begin{equation}
f^{(n)}_\sigma (\vc{r}^n, \vc{v}^n, t) \equiv \frac{N!}{(N-n)!} \int_{\sim V_\sigma} d\vc{\Gamma}^{(N-n)} f^{[N]} (\vc{r}^N, \vc{v}^N,t) .
\end{equation}
Here, $\sim V_\sigma$ denotes that the spatial integral excludes the small volume $V_\sigma$. 
By integrating Liouville's equation~(\ref{eq:liou}), the first-order BBGKY hierarchy equation for the ``truncated reduced distribution function'' is 
\begin{align}
\label{eq:n1_sig_f}
\nonumber
\biggl( \frac{\partial}{\partial t} + \mathcal{L}_1 + \bar{\mathcal{L}}_1^{(1)}& \biggr) f_\sigma^{(1)}(1) = - \oint_{S_2} d \vc{v}_2 d \vc{S}_2 \cdot \vc{u} f_\sigma^{(2)}(1,2) \\ 
& - \int_{\sim V_\sigma} d\vc{r}_2 \int d\vc{v}_2 \mathcal{L}_{12}^{C} \Delta f^{(2)}(1,2) 
\end{align}
where $\vc{u} \equiv \vc{v}_1 - \vc{v}_2$.
Here, an extra term arises in comparison to Eq.~(\ref{eq:n1_f}) that depends on the two-body distribution function evaluated at the surface of the collision volume, $\vc{S}_2$. 
The method relies on the assumption that the collision volume, $V_\sigma$, is sufficiently small that there is an appropriate limit in which $f_\sigma^{(n)} \rightarrow f^{(n)}$ (i.e., that $f_\sigma^{(n)}$ is the observable of interest). 

A key question is, what determines the scale of $V_\sigma$?  
For a dilute gas, this is typically associated with the range of the short-range intermolecular forces. 
In these theories, $f^{(2)}$ replaces $\Delta f^{(2)}$ in Eq.~(\ref{eq:n1_sig_f}) and the last term is considered negligible because the intermolecular force (inside the $\mathcal{L}_{12}$ operator) is small in the region of space outside of the collision volume. 
This term is dropped, and the collision operator is derived by solving the two-body interaction problem inside the collision volume to determine $f^{(2)}$ on its surface. 
This leads to the Boltzmann equation.~\cite{grad:58} 

Clearly, the same argument does not apply to plasmas because the Coulomb force is long-range and effectively extends over all of space. 
Indeed, this is why application of the Boltzmann equation to plasma results in a divergence. 
However, writing Eq.~(\ref{eq:n1_sig_f}) in terms of $\Delta f^{(2)}$ introduces the notion of screening because the effective range of interaction is associated with the spatial correlation scale of $\Delta f^{(2)}$ itself. 
Of course $\Delta f^{(2)}$ vanishes at equilibrium at all spatial scales, so the relevant correlation arises only away from equilibrium. 
Considering a small perturbation from equilibrium, $\Delta f^{(2)}$ will always become small on large scales at which two particles become decorrelated. 
Only at sufficiently small spatial scales (i.e., the collision scale $|\vc{r}_1 - \vc{r}_2| < V_\sigma$) will $\Delta f^{(2)}$ be appreciable in magnitude. 
For instance, consider a homogenous plasma.
Starting at large scales where spatial correlations are small $f^{(2)}(1,2) \approx \rho^2 f^{(1)}(\vc{v}_1) f^{(2)}(\vc{v}_2)$, the spatial component of $\Delta f^{(2)}$ will have the possibility of contributing at the scale associated with spatial correlation of  $f_o^{(2)}$; i.e., $\Delta f^{(2)} \sim \rho^2 - \rho^{(2)} \approx \rho^2 [1 - g(r)]$. 
This scale is directly determined by the potential of mean force via Eq.~(\ref{eq:pomf}), and thus the screening length. 
For example, in a weakly coupled plasma the potential of mean force is the Debye-H\"{u}ckel potential~\cite{deby:23}  
\begin{equation}
w^{(2)}_{\textrm{DH}}(r) = \frac{\phi(r)}{k_BT} e^{-r/\lambda_D}
\end{equation}
where $\lambda_D$ is the Debye length. 
This corresponds directly to the conventional notion that $V_\sigma$ is a ``Debye interaction sphere'' at weak coupling.  
However, the potential of mean force is a more general concept that extends to strongly coupled plasmas as well.

Since $V_\sigma$ is defined as the excluded region of the volume integral, the last term in Eq.~(\ref{eq:n1_sig_f}) is negligible and the kinetic equation is 
\begin{equation}
\label{eq:pke_s}
\biggl( \frac{\partial}{\partial t} + \mathcal{L}_1 + \bar{\mathcal{L}}_1^{(1)} \biggr) f^{(1)}(1) = - \oint_{S_2} d \vc{v}_2 d \vc{S}_2 \cdot \vc{u} f^{(2)}(1,2) .
\end{equation}
Solving for the collision operator in Eq.~(\ref{eq:pke_s}) requires solving for $f^{(2)}$ on the surface of the collision volume. 
As discussed in the previous section, this will be determined from the second order equation [$n=2$ in Eq.~(\ref{eq:bbgky_2})] using $\Delta f^{(3)}=0$ as a closure
\begin{equation}
\label{eq:ke_f2}
\biggl( \frac{\partial}{\partial t} + \mathcal{L}_1 + \mathcal{L}_2 + \bar{\mathcal{L}}_1^{(2)} +  \bar{\mathcal{L}}_2^{(2)} \biggr) f^{(2)}(1,2) = 0  .
\end{equation} 
As does the derivation of the Boltzmann equation, Eq.~(\ref{eq:ke_f2}) describes the dynamical evolution of two interacting particles. 
However, retention of the higher order terms arising in the operators $\bar{\mathcal{L}}_i^{(n)}$ shows that the particles effectively interact via the mean force, rather than the bare Coulomb force $\vc{F}_{ij}^C$. 
In this way, surrounding particles ``mediate'' the binary interactions; a concept Rostoker referred to as ``dielectric dressing''.~\cite{rost:64} 
Thus, although the dynamics are two-body, the interaction is more general than that used in the Boltzmann equation. 
In this sense, it relaxes the binary collision assumption to a certain degree by including the statistical influence of all other $N-2$ particles.

\subsection{Collision Operator\label{sec:wm}} 

The vast majority of plasmas are magnetized in the sense that the Lorentz force may influence macroscopic dynamics, but it does not influence microscopic dynamics within the collision volume.~\cite{baal:17} 
In a weakly coupled plasma, this condition is satisfied if the gyroradius of most particles is signficantly larger than the Debye length: $r_c\gg \lambda_D$, where $r_c = v_T/\omega_c$ is the thermal gyroradius and $\omega_c = |Ze|B/m$ is the gyrofrequency. 
In a strongly coupled plasma, it requires $r_c \gg a$.~\cite{baal:17}  
Most of plasma kinetic theory is based on this scale separation, and this section concentrates on this situation. 
In this case, the Lorentz force term in $\mathcal{L}_1$ contributes to the plasma kinetic equation (\ref{eq:n1_f}), but it is negligible in the operators $\mathcal{L}_1$ and $\mathcal{L}_2$ in Eq.~(\ref{eq:ke_f2}), which describes $f^{(2)}$ inside the collision volume. 

Following the method of Ref.~\onlinecite{grad:58}, we choose a coordinate system such that a diametral plane intersecting the sphere is aligned perpendicular to the relative velocity vector $\vc{u}$. 
Using polar coordinates in this plane, an area element is denoted $d\omega = r dr d\epsilon$. 
This transformation maps the projection of $\vc{S}_2$ onto the disk $0\leq r \leq \sigma$ and $0\leq \epsilon \leq 2\pi$. 
The disk $\omega$ is covered twice: once by the hemisphere $S_2^+:\, \vc{u} \cdot d\vc{S}_2 > 0$ and once by $S_2^-:\, \vc{u} \cdot d\vc{S}_2 < 0$. 
Points that map onto $S_2^+$ represent particles moving away from one another (post-collision particles), while those that map onto $S_2^-$ represent particles moving toward one another (pre-collision particles). 
Denoting the points that map onto $S_2^+$ as $\vc{r}_2^+$ and those that map onto $S_2^-$ as $\vc{r}_2^-$, we find $\vc{u} \cdot d\vc{S}_2 = - u d\omega$ on $S_2^+$ and $\vc{u} \cdot d\vc{S}_2 = u d\omega$ on $S_2^-$. 
In terms of this transformation, Eq.~(\ref{eq:pke_s}) becomes 
\begin{align}
\label{eq:pke_2}
& \biggl( \frac{\partial}{\partial t} + \mathcal{L}_1 + \bar{\mathcal{L}}_1^{(1)} \biggr) f^{(1)}(1) \\ \nonumber
&= \int d \omega d \vc{v}_2 u [f^{(2)}(\vc{r}_1, \vc{v}_1, \vc{r}_2^+,\vc{v}_2) - f^{(2)}(\vc{r}_1, \vc{v}_1, \vc{r}_2^-,\vc{v}_2)].
\end{align}

To obtain an explicit collision operator, we invoke four assumptions that are similar to those underlying the Boltzmann equation: (1) truncation (finite size of $V_\sigma$), (2) binary collisions, (3) modified molecular chaos, and (4) slow variation of $f^{(1)}$. 
Assumption (1) was already applied above, which justified the neglect of the last term in Eq.~(\ref{eq:n1_sig_f}), and that $f_\sigma^{(1)}$ and $f_\sigma^{(2)}$ accurately approximate $f^{(1)}$ and $f^{(2)}$. 
We point out again that assumption (1) has a different origin in the context of the theory presented here, in comparison to the Boltzmann equation. 
Here, the small size of $V_\sigma$ is justified only by the limited range of the effective interaction implied by $\Delta f^{(2)}$. 
In contrast, in terms of $f^{(2)}$ the neglected term in Eq.~(\ref{eq:n1_sig_f}) would diverge for a plasma. 
Neglecting this is justified in derivations of the Boltzmann equation because they focus on dilute gases which have short-range potentials. 

Since the collision volume is small, the number of particle inside of it is very small compared to the total number of particles. 
Furthermore, we assume that within the collision volume particle interactions are predominately binary [assumption (2)], but that they occur via the mean force. 
This justifies $\Delta f^{(3)}=0$ as a closure, leading to Eq.~(\ref{eq:ke_f2}) within $V_\sigma$. 
The solution of this equation is $f^{(2)}=\textrm{const}$ on a 2-particle trajectory. 
Thus, the points that map into the hemisphere $S_2^+$ can be equated with post-collision coordinates: $f^{(2)}(\vc{r}_1, \vc{v}_1, \vc{r}_2^+, \vc{v}_2, t) \rightarrow f^{(2)}(\hat{\vc{r}}_1, \hat{\vc{v}}_1, \hat{\vc{r}}_2, \hat{\vc{v}}_2, \hat{t})$. 

To solve for the particle trajectories inside the sphere, the Boltzmann equation typically makes the ``molecular chaos'' approximation (3) whereby the two-particle distribution is assumed to be uncorrelated in both the initial and final states: $f^{(2)}(1,2) = f^{(1)}(1) f^{(1)}(2)$. 
This is a rather more subtle point associated with the introduction of irreversibility into the theory, which is discussed in depth in Ref.~\onlinecite{grad:58}. 
The procedure presented here attempts to develop a model that is consistent with the equilibrium state. 
As such, we modify this approximation slightly to account for the statistical spatial correlation of particles at the surface of the collision volume: $f^{(2)}(1,2)|_{r=\sigma} \approx \chi f^{(1)}(1) f^{(1)}(2)$. 
This concept, first introduced by Enskog~\cite{ensk:22} in the context of a hard-sphere gas, accounts for the ``excluded volume'' associated with the fact that hard spheres cannot overlap. 
For hard spheres, $\chi = g(r=\sigma)$. 
For a soft potential, $\sigma$ is not a single value for all interactions, and as such the concept is understood as a statistical analog.~\cite{hanl:72} 
This is discussed in Ref.~\onlinecite{baal:15} for single-component plasmas. 
In this model, $\sigma$ is associated with the location where $g(r=\sigma)=0.87$ and $\chi$ is determined by using this effective diameter in the virial expansion of the Enskog equation of state; see also Ref.~\onlinecite{shaf:17} for an extension to binary mixtures. 
The value $0.87$ comes from a fit to molecular dynamics simulations of the OCP, but is expected to be a universal value for repulsive potentials.~\cite{baal:15,dali:16} 
From this model $\chi \rightarrow 1$ for weakly coupled plasmas $\Gamma \lesssim 1$, and $\chi \approx 1.3-1.4$ for $1 \lesssim \Gamma \lesssim 30$. 

Finally, we invoke approximation (4), which states that the collision time is much shorter than the timescale associated with the evolution of $f^{(1)}$. 
In this limit, we expect the spatial coordinates $\hat{\vc{r}}_1$, $\vc{r}_2$ and $\hat{\vc{r}}_2$ to differ from $\vc{r}_1$ by a small amount on the order of $\sigma$. 
Likewise, the short collision time implies that $\hat{t}$ differs from $t$ by a negligible amount. 
Putting the results of these four approximations into Eq.~(\ref{eq:pke_2}) leads to the plasma kinetic equation 
\begin{align}
\label{eq:pke_final}
\biggl( \frac{\partial}{\partial t} &+ \mathcal{L}_1 + \bar{\mathcal{L}}_1^{(1)} \biggr) f^{(1)}(\vc{v}_1) \\ \nonumber
&= \chi \int d \omega d \vc{v}_2 u [f^{(1)}(\hat{\vc{v}}_1)f^{(1)}(\hat{\vc{v}}_2) - f^{(1)}(\vc{v}_1)f^{(1)}(\vc{v}_2) ],
\end{align}
where it is understood that each occurrence  of $f^{(1)}$ is evaluated at the same local position $\vc{r}_1$ and time $t$. 
Equation~(\ref{eq:pke_final}) was derived for a one-component plasma. 
The generalization to a multicomponent system 
\begin{align}
\label{eq:pke_ms}
&\biggl( \frac{\partial}{\partial t} + \mathcal{L}_1 + \bar{\mathcal{L}}_1^{(1)} \biggr) f_s^{(1)}(\vc{v}_1) \\ \nonumber
&= \sum_{s^\prime} \chi_{ss^\prime} \int d \omega d \vc{v}_2 u [f_s^{(1)}(\hat{\vc{v}}_1)f_{s^\prime}^{(1)}(\hat{\vc{v}}_2) - f_s^{(1)}(\vc{v}_1)f_{s^\prime}^{(1)}(\vc{v}_2) ],
\end{align}
follows directly by tracking the derivation above and distinguishing species $s$ from $s^\prime$. 

Since the mean force is central and conservative, the differential area of the disk can be related to a scattering cross section, $d\omega = \sigma_{ss^\prime}\, d\Omega$ where $d\Omega = d\phi d\theta \sin \theta$, or an impact parameter, $d\omega = bdb d\phi$, in the usual way.~\cite{grad:58} 
The results of the standard two-body scattering problem are determined by the scattering angle $\theta = \pi - 2\Theta$, where
\begin{equation}
\Theta = b \int_{r_o}^\infty dr\, r^{-2} \biggl[1 - \frac{b^2}{r^2} - \frac{2 w_{ss^\prime}^{(2)}(r)}{m_{ss^\prime}u^2} \biggr]^{-1/2} .
\end{equation} 
Here, $m_{ss^\prime} = m_sm_{s^\prime}/(m_s+m_{s^\prime})$ is the reduced mass, and $r_o$ is the distance of closest approach, which is determined by the largest root of the term in square brackets. 
Note that this is the same as the scattering angle used in the Boltzmann equation, except that the potential of mean force $w^{(2)}$ replaces the bare interaction potential $\phi(r)$.

\subsection{Comment on Strongly Magnetized Plasmas\label{sec:sm}} 

If the external magnetic field is sufficiently strong that the gyromotion of particles fits within the collision volume, the situation is much more complicated. 
O'Neil considered very strongly magnetized one-component plasmas (such that $r_c \ll r_o$), and derived a kinetic equation by aligning the surface $d\vc{S}_2$ along the magnetic field (rather than $\vc{u}$).~\cite{onei:83} 
A collision operator was then derived based on an assumption that the adiabatic invariant $|u_+|^2/B$ is preserved during the interaction. 
Here $u_+ \equiv u_x+iu_y$ are Cartesian components of $\vc{u}$ and the magnetic field is in the $\hat{z}$ direction. 
This adiabatic assumption is restrictive, and a more general theory is desirable. 

The method outlined above may provide a fruitful starting point for developing such a theory. 
The main advantage is that it self-consistently accounts for screening. 
A matter of continuing investigation in strongly magnetized plasmas is the relative importance of short and long range collisions.~\cite{dubi:14} 
Early theories modified Landau-Spitzer based approaches by changing the Debye length to the gyroradius in the Coulomb logarithm.~\cite{sili:63,mont:74}  
Others modified the Lenard-Balescu equation to account for magnetization in the plasma dielectric function.~\cite{rost:60b,hass:77} 
However, recent molecular dynamics results have shown trends inconsistent with either prediction at strong magnetization.~\cite{baal:17} 
By accounting for screening via the potential of mean force, and for gyromotion via the 2-body interaction inside the collision sphere, one may address in a self-contained way the combined influence of short and long-range interactions. 
The Bohr-van Leeuwen theorem ensures that the magnetic field does not influence any statistical property at equilibrium, so the potential of mean force is expected to remain unchanged from that presented above.  

\subsection{Equation of State\label{sec:eos}} 

In addition to providing a convergent collision operator, Eq.~(\ref{eq:pke_s}) introduces a term in the convective derivative ($\bar{\mathcal{L}}_1^{(1)}$) that is absent from the Boltzmann equation. 
This term is associated with non-ideal components of the equation of state, i.e. the excess (or potential) component of the pressure and internal energy. 
Although this is negligible in a weakly coupled plasma, it is the dominant component in moderately and strongly coupled plasmas.~\cite{hans:06} 
The fact that these terms arise naturally should be expected because the closure is designed to ensure that the exact equilibrium state consistent with the YBG hierarchy, Eq.~(\ref{eq:ybg}), is maintained. 

To make the connection with pressure, it is first useful to notice that the mean force acting on one particle can be written as 
\begin{equation}
\label{eq:f_P}
\bar{\vc{F}}_1^{(1)} =  - \frac{\nabla_1 \cdot \mathcal{P}_\Phi}{\rho^{(1)}(\vc{r}_1)}
\end{equation}
where 
\begin{equation}
\mathcal{P}_\Phi = - \frac{1}{2} \int d\vc{r}\, \vc{r} \vc{r} \frac{\phi^\prime(r)}{r} \int_0^1d \mu\, \rho^{(2)}(\vc{r}_1 - (1-\mu) \vc{r}, \vc{r}_1 + \mu \vc{r}) 
\end{equation}
is a second-rank tensor, $\vc{r} \equiv \vc{r}_2 - \vc{r}_1$ and $\phi^\prime(r) \equiv d\phi/dr$. 
A derivation of Eq.~(\ref{eq:f_P}) from (\ref{eq:Fmf}) is provided in Appendix \ref{appendix}. 
In the common limit that the plasma is sufficiently homogeneous that $\rho^{(1)}(\vc{r}_1)$ is constant on the spatial scale of the two-body correlations, then $\rho^{(2)}(\vc{r}_1, \vc{r}_2) \approx  [\rho^{(1)}(\vc{r}_1)]^2 g(\vc{r}_1, r)$; see Pohl, \text{et al} \cite{pohl:04} for a related discussion.
In this case, the divergence of the pressure tensor in Eq.~(\ref{eq:f_P}) reduces to the gradient of a scalar, $\nabla_1 p_\Phi$, where
\begin{equation}
\label{eq:p_phi_g}
p_\Phi = - \frac{\rho^2}{6} \int_0^\infty d\vc{r}\, \phi^\prime (r) \, r g(r)  
\end{equation}
is the well-known equilibrium expression for the potential component of the scalar pressure.~\cite{hans:06} 

We note that Eq.~(\ref{eq:p_phi_g}) makes use of the assumption of ``local thermodynamic equilibrium'' whereby there is a scale separation between large (fluid-scale) gradients represented by $\vc{r}_1$ and the interaction scale represented by $r = |\vc{r}_1 - \vc{r}_2|$. 
Thus, the force represented by $\nabla_1 p_\Phi$ arises due to a slight inhomogeneity of the background density across the collision volume ($L \gg \sigma$). 
This origin of the excess pressure can be compared with Enskog's theory of hard sphere gases.~\cite{ensk:22,ferz:72} 
Enskog's theory expands $f^{(1)}(\vc{r}_1\pm \sigma_\textrm{s} \vc{k})$ about the local position $\vc{r}_1$, where $\sigma_s$ is the sphere diameter and $\vc{k}$ is vector connecting the center of two spheres at the instant of contact. 
The excess pressure (and excess internal energy) arising from the lowest-order term in this expansion comes about due to any inhomogeneity of $f^{(1)}(\vc{r}_1)$ across the collision volume, which is simply the volume of the sphere in a hard-sphere gas. 
The higher order terms proportional to $\nabla_1 f^{(1)}$, etc., are associated with the non-local nature of the collision itself (a hard sphere gas is a singular case in that $\sigma_\textrm{s}$ is both the maximum and minimum scale of an interaction). 
Equation~(\ref{eq:p_phi_g}) represents a generalization of the first of these contributions (local collisions taking place in a finite sized collision volume) to the case of arbitrary potential. 
Its existence in the theory is enforced by the requirement that the exact equilibrium, or local equilibrium, limit be maintained. 

A closed fluid description follows from a Chapman-Enskog solution of the kinetic equation,~\cite{ferz:72} but the basic features of non-ideality can be illustrated directly from the conservation equations. 
In particular, the density moment of Eq.~(\ref{eq:pke_ms}) leads to the usual continuity equation 
\begin{equation}
\frac{\partial \rho}{\partial t} + \nabla \cdot (\rho \vc{V}) = 0
\end{equation}
where $\vc{V} \equiv \int d\vc{v}\, \vc{v} f^{(1)}/\rho$ is the hydrodynamic velocity and $\rho = \rho^{(1)}(\vc{r}_1)$. 
This is the same as what results from the Boltzmann equation. 
However, the momentum moment leads to
\begin{equation}
\rho_m \frac{d\vc{V}}{dt} = - \nabla \cdot \mathcal{P} + \rho_q \vc{E}^\prime + \vc{J} \times \vc{B}
\end{equation} 
in which the total pressure tensor now includes both kinetic (ideal) and potential (excess) components  
\begin{equation}
\mathcal{P} = \mathcal{P}_K + \mathcal{P}_\Phi .
\end{equation} 
Here, $d/dt = \partial/\partial t + \vc{V} \cdot \nabla$, $\mathcal{P}_K = \sum_s m_s \int d\vc{v}\, \vc{v}_r \vc{v}_r f_s^{(1)}/\rho$ is the kinetic part of the pressure tensor, $\vc{v}_r = \vc{v} - \vc{V}$, $\rho_m = \sum_s m_s \rho_s$ is the mass density, $\rho_q = \sum_s q_s \rho_s$ the charge density, $\vc{J} = \sum_s q_s \rho_s (\vc{V}_s - \vc{V})$ the current density, and $\vc{E}^\prime = \vc{E} + \vc{V} \times \vc{B}$ the electric field in the reference frame of the fluid. 

The mean force will also contribute to the energy balance. 
Taking the $\frac{1}{2} m_s \vc{v}_{r}^2$ moment of the kinetic equation, and summing over species, gives 
\begin{equation}
\label{eq:uk_1}
\rho \frac{du_K}{dt} = - \mathcal{P}_K \colon \nabla \vc{V} - \nabla \cdot \vc{q}_K + \vc{J} \cdot \vc{E}^\prime - \vc{V} \cdot (\nabla \cdot \mathcal{P}_\Phi) ,
\end{equation} 
where $u_K = \sum_s \frac{1}{2} m_s \int d\vc{v}\, v_r^2 f_s^{(1)}$ is the internal kinetic energy, and $\vc{q} = \sum_s \frac{1}{2}m_s \int d\vc{v}\, \vc{v}_r v_r^2 f_s^{(1)}$ is the kinetic component of the heat flux. 
Noting that $\vc{V} \cdot (\nabla \cdot \mathcal{P}_\Phi) = \nabla \cdot (\vc{V} \cdot \mathcal{P}_\Phi) - \mathcal{P}_\Phi \colon \nabla \vc{V}$, Eq.~(\ref{eq:uk_1}) can alternatively be written  
\begin{equation}
\rho \frac{du_K}{dt} = - \mathcal{P} \colon \nabla \vc{V} - \nabla \cdot \vc{q} + \vc{J} \cdot \vc{E}^\prime,
\end{equation}
where $\vc{q} = \vc{q}_K + \vc{q}_{\Phi}$, and \
\begin{equation}
\vc{q}_{\Phi} = \vc{V} \cdot \mathcal{P}_\Phi
\end{equation}
is a contribution to the heat flux from the mean force that is absent in the Boltzmann equation. 

An evolution equation for the total internal energy must also include the contribution from the average interparticle potential energy 
\begin{equation}
\label{eq:u_phi}
\rho u_\Phi = \frac{1}{2} \int \phi (r) f^{(2)}(1,2;t)\, d\vc{v}_1 d\vc{v}_2 d\vc{r}_2|_{\vc{r}_1=\vc{r}} .
\end{equation} 
The evolution of $u_\Phi$ is described by the $\frac{1}{2}\phi(r)$ moment of the second order equation of the BBGKY hierarchy 
\begin{align}
\label{eq:u_phi_1}
\rho & \frac{d u_\Phi}{dt} = - \frac{1}{2} \nabla_{\vc{r}_1} \cdot \int \phi(r) \vc{v}_{r1}  f^{(2)}(1,2;t) d\vc{v}_1 d\vc{v}_2 d\vc{r}_2|_{\vc{r}_1=\vc{r}} \nonumber \\ 
&- \frac{1}{2} \int \vc{u} \cdot (\nabla_{\vc{r}_1} \phi) f^{(2)}(1,2;t)d\vc{v}_1 d\vc{v}_2 d\vc{r}_2|_{\vc{r}_1=\vc{r}} .
\end{align}
Explicit evaluation of this requires a time-dependent solution for $f^{(2)}$. 
However, if we apply a Kirkwood-like superposition approximation~\cite{kirk:46} at second order, $f^{(2)}(1,2) \approx \rho^{(2)}(\vc{r}_1,\vc{r}_2) f^{(1)}(\vc{v}_1) f^{(2)}(\vc{v}_2)$, as in Sec.~\ref{sec:wm} above, Eq~(\ref{eq:u_phi}) reduces to $\rho u_\Phi = \frac{1}{2} \int \phi(r) \rho^{(2)}(\vc{r}_1, \vc{r}_2)d\vc{r}_2|_{\vc{r}_1=\vc{r}}$, and Eq.~(\ref{eq:u_phi_1}) to $du_\Phi/dt=0$. 
In this limit, the total energy evolution equation then has the form 
\begin{equation}
\rho \frac{du}{dt} = - \mathcal{P} \colon \nabla \vc{V} - \nabla \cdot \vc{q} + \vc{J} \cdot \vc{E}^\prime ,
\end{equation}
where $u = u_K + u_\Phi$ is the total internal energy. 
Note that because of the non-ideal terms, thermodynamic relations must be invoked to relate this to temperature.~\cite{ferz:72}  
In the limit that the plasma is locally homogeneous, the familiar expression for the potential energy density~\cite{hans:06} 
\begin{equation}
u_\Phi = \frac{\rho}{2} \int d\vc{r}\, \phi(r) g(r) 
\end{equation}
is obtained.

\section{Comparison with previous models\label{sec:comp}} 

Common plasma theories can also be derived from closures of the BBGKY hierarchy. 
Here, these closures are compared with the one presented above. 
Specific attention is paid to the approximations that lead to divergences by considering the equilibrium limit of the second order distribution function resulting from each closure. 

A common first step in the derivation of these equations is to define a cluster expansion of the form 
\begin{subequations}
\label{eq:cluster}
\begin{align}
f^{(1)}(1) &= f(1) \\ 
f^{(2)}(1,2) & = f(1)f(2) + g_2(1,2) \\ 
f^{(3)}(1,2,3) &= f(1) f(2) f(3) + f(1) g_2(2,3) \\ \nonumber
&+ f(2) g_2(1,3) + f(3) g_2(1,2) + g_3(1,2,3)  .
\end{align} 
\end{subequations}
The cluster expansion allows aspects of the triplet distribution function to be retained via products of $f^{(1)}$ and $g_2$. 
Applying this, Eq.~(\ref{eq:f2}) is 
\begin{subequations}
\label{eq:g2}
\begin{align}
\label{eq:ga}
\biggl( & \frac{\partial}{\partial t} + \mathcal{L}_1 + \mathcal{L}_2 \biggr) g_2(1,2) + \bigl( \mathcal{L}_{12}^C + \mathcal{L}_{21}^C \bigr) f(1) f(2)   \\ 
\label{eq:gb}
& = - \bigl(\mathcal{L}_{12}^C + \mathcal{L}_{21}^C \bigr) g_2(1,2) \\ 
\label{eq:gc}
& - \int d\vc{\Gamma}_3 \bigl[ \mathcal{L}_{13}^C g_{2}(2,3)f(1) + \mathcal{L}_{23}^C g_{2}(1,3) f(2) \bigr] \\ 
\label{eq:gd}
& - \int d \vc{\Gamma}_3 \bigl( \mathcal{L}_{13}^C + \mathcal{L}^C_{23} \bigr) g_3(1,2,3)  .
\end{align}
\end{subequations}
Each of the theories discussed below are based on a binary collision
approximation in which the triplet correlation term Eq.~(\ref{eq:gd})
is neglected, but where each keeps a different subset of terms (\ref{eq:ga})--(\ref{eq:gc}) selected, as illustrated below for the Landau and Boltzmann equations, by identifying the leading order in the expansion parameters discussed in the introduction.\cite{Balescu_book}
 
Divergences arise in each, not because of the neglect of triplet correlations, but rather because of neglecting one or more of the terms (\ref{eq:gb})--(\ref{eq:gc}). 
Although the collision operator vanishes at equilibrium due to the velocity dependence of the Maxwellian distribution, the divergences are associated with the spatial component of the distribution function. 
Thus, they can be illustrated by considering just the spatial dependence of the collision operator at equilibrium 
\begin{equation}
\label{eq:cpto}
C \propto \int d\vc{r}_2 \mathcal{L}_{12}^C g_2^{\textrm{eq}}(1,2) \propto \int dr [1 - g(r)]
\end{equation}
where  
\begin{equation}
\label{eq:g2_eq}
g_2^{\textrm{eq}}(1,2) = \rho^2 f_\textrm{M}(\vc{v}_1) f_\textrm{M}(\vc{v}_2) [ g(r) - 1] .  
\end{equation}

\emph{Landau Equation}: 
As recalled in the introduction, the Landau equation is obtained in the weakly coupled limit characterized by $\lambda=\phi_0/k_\bb T\ll 1$.
In terms of this perturbative parameter, the Liouville operators satisfy $\mathcal{L}_i=\mathcal{O}(\lambda^0)$ and $\mathcal{L}_{ij}=\mathcal{O}(\lambda^1)$, while the distribution functions $f=\mathcal{O}(\lambda^0)$ (since it carries the complete normalization regardless of $\lambda$) and it is assumed $g_n=\mathcal{O}(\lambda^{n-1})$ (e.g., at least one direct interaction is required to create a two-body correlation from an uncorrelated state).
Using these orderings in Eq.~(\ref{eq:g2}) and keeping the lowest order in $\lambda$, Landau's seminal kinetic equation~\cite{land:36} can be derived from Eq.~(\ref{eq:g2}) by keeping only term (\ref{eq:ga}) to model $g_2$. 
Solving Eq.~(\ref{eq:ga}) at equilibrium implies  
\begin{equation} 
g(r) = 1 - \frac{\phi(r)}{k_\bb T} .
\end{equation} 
Using this result in Eq.~(\ref{eq:cpto}) shows that 
\begin{equation}
C_\textrm{L} \propto \int dr \frac{\phi(r)}{k_\bb T} \propto \int dr\, \frac{1}{r} , 
\end{equation}
which diverges logarithmically in both the large and small $r$ limits. 
This indicates that the Landau closure neglects both short range physics [term (\ref{eq:gb})] and screening [term (\ref{eq:gc})].
Analogously, the same logarithmic divergences would be observed in the terms (\ref{eq:gb}) and (\ref{eq:gc}), but these are dropped from the analysis. 
The same result is obtained for other plasma kinetic equations that are equivalent to the Landau equation, such as Rosenbluth's Fokker-Planck equation.~\cite{rose:57} 
Landau argued that the limits of integration should range from the thermal distance of closest approach in a binary collision, $r_{\textrm{L}} = e^2/k_BT$, to the Debye length $\lambda_D$.

\emph{Boltzmann Equation}:
The Boltzmann equation assumes $\lambda=nl^3\ll 1$.
In terms of this perturbative parameter, the Liouville operators satisfy $\mathcal{L}_i=\mathcal{O}(\lambda^0)$ and $\mathcal{L}_{ij}=\mathcal{O}(\lambda^0)$, and, from their dependence on the particle density, the distribution functions $f=\mathcal{O}(\lambda^1)$ and $g_n=\mathcal{O}(\lambda^{n})$.
Using these orderings in Eq.~(\ref{eq:g2}) and keeping the lowest order in $\lambda$, the Boltzmann equation can be derived from Eq.~(\ref{eq:g2}) by keeping terms (\ref{eq:ga}) and (\ref{eq:gb}), while dropping terms (\ref{eq:gc}) and (\ref{eq:gd}) (this is equivalent to taking a closure $f^{(3)}=0$ and applying the method of Sec.~\ref{sec:c_op}).~\cite{ferz:72} 
Substituting Eq.~(\ref{eq:g2_eq}) into the equation resulting from this approximation shows that it implies
\begin{equation}
\label{eq:g_boltz}
g(r) = e^{- \phi(r)/k_\bb T} 
\end{equation}
at equilibrium. 
Using this result in Eq.~(\ref{eq:cpto}) shows that 
\begin{equation}
C_\textrm{B} \propto \int dr \biggl[ \exp\biggl(- \frac{\phi(r)}{k_\bb T} \biggr) - 1 \biggr] .
\end{equation}
This expression converges in the close interaction limit ($r \rightarrow 0$), but  diverges logarithmically in the far interaction limit. 
Analogously, the same logarithmic divergence would be observed in the term Eq.~(\ref{eq:gc}) that is dropped from the analysis. 
Physically, this is because screening is neglected along with Eq.~(\ref{eq:gc}). 
When the Boltzmann equation is used in plasma theory, interactions are limited to within a Debye length $\lambda_D$ \emph{ad hoc}. 

\emph{Lenard-Balescu Equation}: The Lenard-Balescu equation~\cite{lena:60,bale:60,guer:60} can be derived from Eq.~(\ref{eq:g2}) by keeping terms (\ref{eq:ga}) and (\ref{eq:gc}) [while dropping terms (\ref{eq:gb}) and (\ref{eq:gd})].~\cite{nich:83}
Substituting Eq.~(\ref{eq:g2_eq}) into the resultant approximation shows that this implies 
\begin{equation}
g(r) = 1 - \frac{\phi(r)}{k_\bb T} e^{-r/\lambda_D} .
\end{equation} 
Using this result in Eq.~(\ref{eq:cpto}) shows that 
\begin{equation}
C_\textrm{LB} \propto \int dr \frac{\phi(r)}{k_\bb T} e^{-r/\lambda_D} . 
\end{equation} 
This expression converges in the far interaction limit ($r \rightarrow \infty$), but since the integrand is proportional to $1/r$ for small $r$, it diverges logarithmically in the close interaction limit. 
The neglected term (\ref{eq:gb}) contains the physics of the close interaction. 
This limit is typically resolved by truncating the spatial integral at the thermal distance of closest approach in a binary collision, $r_{\textrm{L}}$. 

\emph{Frieman-Book Equation}: Self-consistently convergent kinetic equations have previously been developed. 
Frieman and Book \cite{frie:63} derived such an equation by matching solutions of Eq.~(\ref{eq:g2}) in asymptotic limits of close interaction ($r \lesssim r_\textrm{L}$), far interaction ($r \gtrsim \lambda_D$) and intermediate scale ($r \sim n^{-1/3}$). 
The solution in each limit is obtained by neglecting the terms described above for the Boltzmann equation, Lenard-Balescu equation and Landau equation, respectively. 
The resultant expression for $g_2^{\textrm{eq}}$ consists of the sum of the Boltzmann equation plus the Lenard-Balescu equation minus the Landau equation. 
The radial distribution function then has the form
\begin{equation}
g(r) = e^{-\phi/k_\bb T}  - \frac{\phi}{k_\bb T} e^{-r/\lambda_D} + \frac{\phi}{k_\bb T} 
\end{equation}
(see also Refs.~\onlinecite{shur:64,dewi:65} for further discussion).
Using this result in Eqs.~(\ref{eq:cpto}) and (\ref{eq:g2_eq}) shows that 
\begin{equation}
C_\textrm{FB} \propto \int dr \biggl(1 - e^{-\phi/k_\bb T}  - \frac{\phi}{k_\bb T} e^{-r/\lambda_D} + \frac{\phi}{k_\bb T}\biggr) .
\end{equation}
This expression converges in both the close and far interaction limits. 
This shows that the divergences in the other theories are associated with neglecting one of the terms (\ref{eq:gb})-(\ref{eq:gc}), rather than the triplet correlation (\ref{eq:gd}). 

\emph{Relation with the theory of Sec.~\ref{sec:c_op}}: Since this procedure preserves the exact equilibrium limit, the relationship implied by the second order equation  is simply the exact statistical statement~\cite{hill:60}  
\begin{equation}
g(r) = \exp \biggl[ - \frac{w^{(2)}(r)}{k_\bb T} \biggr]  .
\end{equation}
However, a closure of Eq.~(\ref{eq:ybg}) is still required to determine $g(r)$. 
Since this is an equilibrium quantity, the tools of equilibrium statistical mechanics are available for this task, and accurate approximations have been developed. 
In principle, any such approximation for $g(r)$, or even experimental data, can serve as the input to this theory. 

One example of an accurate closure for Coulomb systems is the combination of the hypernetted chain and Ornstein-Zernike equations~\cite{hans:06}
\begin{subequations}
\label{eq:hnc}
\begin{align}
g(r) &= \exp[-\phi(r)/k_\bb T + h(r) - c(r)] \\
\hat{h}(k) & = \hat{c}(k) [ 1 + n\hat{h}(k)] 
\end{align}
\end{subequations}
where $h(r) \equiv g(r) - 1$, and $\hat{h}(k)$ denotes the Fourier transform of $h(r)$. 
This closed set of equations is known to accurately describe $g(r)$ in plasmas spanning from asymptotically weak coupling well into the strong coupling regime. 
Further extensions have been provided, via models for the bridge function that extend this even further to near solid-state plasma conditions.~\cite{tana:86,ichi:92} 
Thus, very accurate methods are available to determine $g(r)$ at equilibrium at essentially any conditions. 

At weak coupling, Eq.~(\ref{eq:hnc}) reduces to the Debye-H\"{u}ckel limit   
\begin{equation}
\label{eq:gr_dh}
g(r) = \exp \biggl[ - \frac{\phi(r)}{k_\bb T} e^{-r/\lambda_D} \biggr]  .
\end{equation}
O'Neil and Rostoker showed that this is valid to order $\Gamma^3 \ln \Gamma$ for $\Gamma \ll 1$.~\cite{onei:65} 
None of the approximate theories described above capture Eq.~(\ref{eq:gr_dh}). 
However, the convergent Frieman-Book result is consistent with it to order $\Gamma^2$; see Refs.~\onlinecite{shur:64,dewi:65} for further discussion. 

Other previous kinetic theories, including Liboff~\cite{libo:59} and Paquette,~\cite{paqu:86} have modified the Boltzmann equation in an ad-hoc manner by modeling binary collisions as occurring via the Debye-H\"{u}ckel potential (rather than the Coulomb potential). 
This leads to a convergent kinetic equation that accurately models weakly coupled plasmas. 
Indeed, if $\phi(r)$ is replaced by the Debye-H\"{u}ckel potential in Eq.~(\ref{eq:g_boltz}) the correct result for $g_2$ at equilibrium [Eq.~(\ref{eq:gr_dh})] is obtained. 

In a sense, these theories, as well as Lenard-Balescu  theory, can be understood as variants of the ``dressed test particle'' concept.~\cite{rost:64} 
Differences lie in the details of how interactions are modeled. 
Lenard-Balescu type approaches model interactions via the correlation of linear fluctuations. 
This loses the ``particle'' concept, and thus does not include close collisions, but it does account for dynamics in the dielectric ``dressing'', which can be important for super-thermal particles. 
For instance, the dynamic aspect of the dielectric response results in order-unity contributions to fast particle stopping in a plasma.~\cite{zwic:99} 
In contrast, the modified Boltzmann based approaches retain the particle concept, and physics of close collisions. 
However, the dielectric response in this case is input in an ad hoc manner, via the screened Coulomb potential, which is accurate for sub-thermal particles~\cite{nich:83} and thus expected to apply to near equilibrium transport processes.  

The approach of Sec.~\ref{sec:c_op} essentially formalizes the latter line of reasoning by enforcing the viewpoint that the exact equilibrium limit should be maintained in the closure to the BBGKY hierarchy. 
Since this is the Debye-H\"{u}ckel result at weak coupling, the theory reduces to the result of Liboff~\cite{libo:59} in that limit. 
It also extends this line of reasoning by revealing that the appropriate interaction potential is the potential of mean force, which differs substantially from the Debye-H\"{u}ckel potential at strong Coulomb coupling. 
Recent work has shown that modeling the potential of mean force via Eq.~(\ref{eq:hnc}) enables an extension of traditional plasma transport theory well into the strongly coupled regime (typically for $\Gamma \lesssim 20$).~\cite{baal:13,baal:14,baal:15,dali:14}  
Similar results have also been obtained in non-equilibrium two-temperature plasmas.~\cite{seuf:89,shaf:17b,shaf:19}

\section{Conclusions and outlook~\label{sec:conc}} 

The closure presented in this paper makes two advances. 
First, it provides a formal expansion parameter [Eq.~(\ref{eq:dfn})] for the BBGKY hierarchy that enables a self-contained derivation of a plasma kinetic equation that includes both static screening and close interactions. 
Second, it provides a means to extend traditional plasma theory to strong Coulomb coupling, not only in the collision operator, but also in the equation of state. 
It does so by ensuring that the equilibrium limit is preserved in the spatial correlation at second order of the hierarchy. 

This formalizes and extends the EPT collision operator obtained previously from physical arguments.~\cite{baal:13,baal:14,baal:15} 
It also provides a conceptual basis that may prove useful for extending the theory to other regimes of plasma physics. 
For example, kinetic theories have been developed to include three-body interactions.~\cite{onei:65,choh:58} 
A similar analysis could be applied while using $\Delta f^{(4)}=0$ to close the hierarchy, thereby including the static screening response in effective three-body dynamics. 
It may provide a basis for addressing strongly magnetized plasmas in which the Lorentz force acts at the collision scale. 
Finally, the approach was motivated by plasmas, which emphasize the need to account for many-body screening self-consistently. 
However, the theory may also be applied to other types of systems, particularly in the kinetic theory of dense gases.~\cite{ferz:72}

\appendix \section{Derivation of Equation~(\ref{eq:fmf})} 
\label{pomf_appendix} 

Here, it is shown that the mean force given by Eqs.~(\ref{eq:fmf}) and (\ref{eq:pomf}) is equivalent to (\ref{eq:Fmf}). 
First, note that for $i\leq n$,
\begin{eqnarray*}
\lefteqn{-\nabla_i \left[V_{\ext}(\vc{r}^N)+V_N (\vc{r}^N)\right]}&&\\
&=& \vc{F}_i^{\ext} +\sum_{\stackrel{j=1}{j\neq i}}^n \vc{F}_{ij}^C +\sum_{j=n+1}^N \vc{F}_{ij}^C 
\end{eqnarray*}
so that the gradient of the $n-$particle density $\rho^{(n)} (\vc{r}^n)$ defined by Eq.~(\ref{rhonrn}) can be written
\begin{align*}
&k_\bb T\,\nabla_i \rho^{(n)}(\vc{r}^n)\\
&=\frac{N!}{(N-n)!} \frac{1}{\mathcal{Z}_N}\int{dr^{N-n} e^{-(V_{\ext}+V_N)/k_\bb T}}\\
&\quad\quad\quad\quad\quad\times\left[\vc{F}_i^{\ext}(\vc{r}_i) +\sum_{\stackrel{j=1}{j\neq i}}^n \vc{F}_{ij}^C +\sum_{j=n+1}^N \vc{F}_{ij}^C\right]\\
&=\left[\vc{F}_i^{\ext}(\vc{r}_i) +\sum_{\stackrel{j=1}{j\neq i}}^n \vc{F}_{ij}^C \right]\rho^{(n)}(\vc{r}^n)\\
&\quad+\frac{N!}{(N-n)!} \frac{1}{\mathcal{Z}_N}\int{dr^{N-n} \sum_{j=n+1}^N \vc{F}_{ij}^C\,e^{-(V_{\ext}+V_N)/k_\bb T}}\,.
\end{align*}
Using,
\begin{align*}
&\int{dr^{N-n} \sum_{j=n+1}^N \vc{F}_{ij}^C\,e^{-(V_{\ext}+V_N)/k_\bb T}}\\
&=(N-n) \int{dr^{N-n} \vc{F}_{i,n+1}^C\,e^{-(V_{\ext}+V_N)/k_\bb T}}\\
&=(N-n) \mathcal{Z}_N\frac{(N-(n+1))!}{N!}\int{dr^{N-n} \vc{F}_{i,n+1}^C\rho^{(n+1)}}\\
&=\mathcal{Z}_N\frac{(N-n)!}{N!}\int{dr^{N-n} \vc{F}_{i,n+1}^C\rho^{(n+1)}}\,,
\end{align*}
we find
\begin{align*}
&k_\bb T\,\nabla_i \rho^{(n)}(\vc{r}^n)/\rho^{(n)}(\vc{r}^n)\\
&=\vc{F}_i^{\ext}(\vc{r}_i) +\sum_{\stackrel{j=1}{j\neq i}}^n \vc{F}_{ij}^C +\int{dr^{N-n} \vc{F}_{i,n+1}^C\frac{\rho^{(n+1)}}{\rho^{(n)}}}\,,
\end{align*}
which completes the proof. 

\section{Derivation of Equation~(\ref{eq:f_P})} 
\label{appendix}

Equation~(\ref{eq:f_P}) follows from the $n=1$ component of Eq.~(\ref{eq:Fmf})
\begin{equation}
\label{eq:F1_1}
\bar{\vc{F}}_1^{(1)} = - \int d^3r_2 [\nabla_1\phi(r)] \frac{\rho^{(2)}(\vc{r}_1, \vc{r}_2)}{\rho^{(1)}(\vc{r}_1)} .
\end{equation}
This derivation follows an analogous method presented in Ref.~\onlinecite{ferz:72}.
First, note that  
\begin{subequations}
\begin{align}
\int d^3r_2 \nabla_1 \phi(r) \rho^{(2)}(\vc{r}_1, \vc{r}_2) &= - \int d^3r\, \phi^\prime(r) \frac{\vc{r}}{r} \rho^{(2)}(\vc{r}_1, \vc{r} + \vc{r}_1) \\
\label{eq:n2_b}
&= - \int d^3r\, \phi^\prime(r) \frac{\vc{r}}{r} \rho^{(2)}(\vc{r}_1 + \vc{r}, \vc{r}_1) 
\end{align}
\end{subequations}
where the second line follows from the property that $\rho^{(2)}$ is constant under the interchange $1 \leftrightarrow 2$. 
Combining these two equivalent expressions, and applying the substitution $\vc{r} \rightarrow - \vc{r}$ as the integration variable in Eq.~(\ref{eq:n2_b}), gives 
\begin{align}
\label{eq:phi_mid}
&\int d^3r_2 \nabla_1 \phi(r) \rho^{(2)}(\vc{r}_1, \vc{r}_2) = \\ \nonumber
 &-\frac{1}{2} \int d^3r\, \phi^\prime(r) \frac{\vc{r}}{r} [\rho^{(2)}(\vc{r}_1, \vc{r}_1 + \vc{r}) - \rho^{(2)}(\vc{r}_1 - \vc{r}, \vc{r}_1)] .
\end{align} 
Next, observing that 
\begin{align}
\nonumber
&\rho^{(2)}(\vc{r}_1, \vc{r}_1 + \vc{r}) - \rho^{(2)}(\vc{r}_1 - \vc{r}, \vc{r}_1) \\
&= \int_0^1 d\mu\, \frac{\partial}{\partial \mu} \rho^{(2)}(\vc{r}_1 - (1-\mu)\vc{r}, \vc{r}_1 + \mu \vc{r}) \\
&= \int_0^1 d\mu\, \vc{r} \cdot \nabla_1 \rho^{(2)}(\vc{r}_1 - (1-\mu)\vc{r}, \vc{r}_1 + \mu \vc{r})
\end{align}
shows that Eq.~(\ref{eq:phi_mid}) can be written as the divergence of a tensor 
\begin{align}
\label{eq:phi_end}
&\int d^3r_2 [\nabla_1 \phi(r)] \rho^{(2)}(\vc{r}_1, \vc{r}_2) = \\ \nonumber
&- \nabla_1 \cdot \frac{1}{2} \int d^3r\, \vc{r} \vc{r} \frac{\phi^\prime(r)}{r} \int_0^1 d\mu \rho^{(2)}(\vc{r}_1 - (1-\mu) \vc{r}, \vc{r}_1 + \mu \vc{r}) .
\end{align}
Putting Eq.~(\ref{eq:phi_end}) into (\ref{eq:F1_1}) completes the derivation of Eq.~(\ref{eq:f_P}).

\begin{acknowledgments}

The authors thank Nathaniel Shaffer and Louis Jose for helpful discussions. 
This work was supported by the U.~S.~Air Force Office of Scientific Research under Award No.~FA9550-16-1-0221; and by the U.~S.~Department of Energy, Office of Fusion Energy Sciences, under Award No.~DE-SC0016159. 
The work of J.~D.~was supported by the US Department of Energy through the Los Alamos National Laboratory. Los Alamos National Laboratory is operated by Triad National Security, LLC, for the National Nuclear Security Administration of U.~S.~Department of Energy (Contract No. 89233218CNA000001).


\end{acknowledgments}


\end{document}